\documentclass[fleqn,twoside]{article}
\usepackage{espcrc2}

\usepackage{graphicx}
\usepackage[figuresright]{rotating}

\newcommand{\AmS}{{\protect\the\textfont2
  A\kern-.1667em\lower.5ex\hbox{M}\kern-.125emS}}

\title{$D_s$ spectrum and leptonic decays
with Fermilab heavy quarks
 and improved staggered light quarks
}

\author{Massimo~di~Pierro\address[DePaul]{School of Computer 
Science, Telecommunications, and Information Systems, DePaul University, Chicago, IL 60604},
Aida~X.~El-Khadra\address[UIUC]{
Department of Physics, University of Illinois, Urbana, IL 61801},
Steven~Gottlieb\addressmark[FNAL]\address[IU]{
Department of Physics, Indiana University, Bloomington, IN 47405},
Andreas~S.~Kronfeld\address[FNAL]{
Fermi National Accelerator Laboratory, P.O. Box 500, Batavia, IL 60510}, 
Paul~B.~Mackenzie\addressmark[FNAL]\thanks{Talk given by Paul Mackenzie, mackenzie@fnal.gov},
Damian~P.~Menscher\addressmark[UIUC], Mehmet~B.~Oktay\addressmark[UIUC],
Masataka Okamoto\addressmark[FNAL],
and James~N.~Simone\addressmark[FNAL]
}

\begin{document}

\begin{abstract}
We present preliminary results for the $D_s$ meson spectrum and decay constants
in unquenched lattice QCD.
Simulations are carried out with $2+1$ dynamical quarks using gauge
configurations generated by the MILC collaboration.
We use the ``asqtad'' $a^2$ improved staggered action for the light
quarks, and the clover heavy quark action with the Fermilab
interpretation.
We compare our spectrum results with the newly discovered $0^+$
and $1^+$ states  in the $D_s$ system.
\vspace{1pc}
\end{abstract}

\maketitle

\section{INTRODUCTION}

$D$ physics is assuming a larger importance in lattice QCD and
in CKM phenomenology with the arrival of the CLEO-c
charm factory \cite{Briere:2001rn}.
CLEO-c  will measure the leptonic decay constants 
$f_D$ and $f_{D_s}$ to an accuracy of around 2\%.
It will measure the amplitudes of the semileptonic decays
$D\rightarrow \pi l \nu$  and $D\rightarrow K l \nu$  to an
accuracy of around 1\%.
This will produce new determinations of the CKM matrix elements
$V_{cd}$ 
and $V_{cs}$  to the accuracy that
can be achieved in the required lattice calculations,
and  new checks of the unitarity triangle.
New, precise tests of lattice QCD will come from the amplitude
ratios $f_D/D\rightarrow \pi l \nu$ and
$f_{D_s}/D\rightarrow K l \nu$.
These ratios, which are independent of the CKM matrix, will provide
the most precise tests in existence of the types of
 lattice QCD calculations required to extract CKM matrix elements
from $B$ physics and $D$ physics.

The recent discovery of the positive parity partners
of the $D_s$ and the $D_s^*$ \cite{Besson:2003cp}
 has added new interest to the spectrum of the $D_s$ system.
The new states lie significantly below quark model predictions,
 and below the $D K$ threshold, so the states are quite narrow.

\section{METHODS}

We use the ``asqtad'' order $a^2$ improved staggered light quark action,
and the order $a$ improved Fermilab heavy quark action.
In unquenched calculations, chiral extrapolation is the least well
controlled remaining source of error.
staggered fermions are the only method currently able to reach the 
region $m_l\sim m_s/5$ that seems to be required to control this 
error.
Using unquenched calculations that reach this region,
a number of simple quantities that disagree with each other at
the 10\% level in the quenched approximation come into good 
agreement \cite{Davies:2003ik}.
We use the public MILC unquenched configurations, with
two light and one strange sea quarks, with properties
\begin{itemize}
  \item $20^3\times 64$, $a\sim$1/8 fm,
  \item ($m_l,m_s$) = (0.007,0.05), (0.01,0.05), (0.02,0.05), (0.03,0.05),
     (True $m_s \approx$ 0.041.),
  \item $\sim$ 500 configurations at each mass, 4 time sources each,
\end{itemize}
as described in \cite{Bernard:2001av}.
\begin{figure}[htb]
\vspace{9pt}
\rotatebox{270}{\includegraphics[width=55mm]{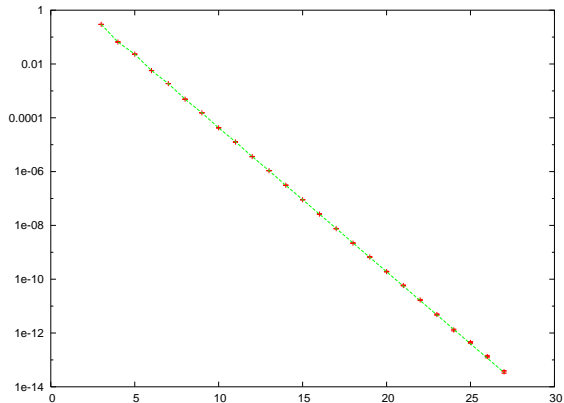}}
\caption{The $D_s$ correlation function.
}
\label{fig:plotg}
\end{figure}
\begin{figure}[htb]
\vspace{9pt}
\rotatebox{270}{\includegraphics[width=55mm]{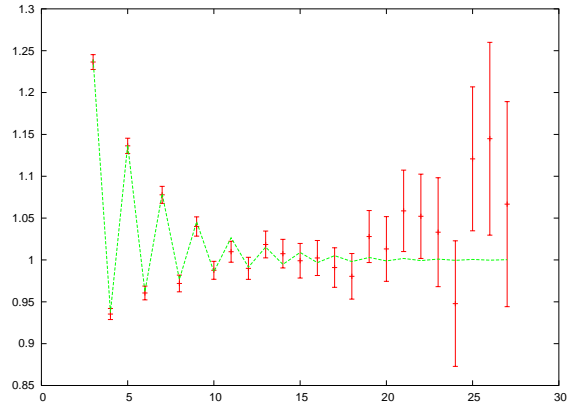}}
\caption{the $D_s$ correlation function with the leading exponential
scaled out.
The exponential decay of the oscillating signal gives the mass
of the $0^+$ state.
}
\label{fig:plotf}
\end{figure}

With staggered and naive light quarks,
the $0^+$ and $1^+$ are  in the $D_s$ and $D_s^*$
correlators automatically and
unavoidably because of fermion doubling.
Naive fermions possess a doubling symmetry,
$\psi\rightarrow (i\gamma_\mu \gamma_5)(-1)^\mu\psi
$,
that implies that quarks come in multiples with identical properties.
Local fermion operators connect to all of these doubled modes.
Therefore, because of the doubling symmetry,
an pseudoscalar current couples to a positive parity scalar state as
well:
$\Psi_h \gamma_5 \psi \rightarrow \Psi_h \gamma_5(i\gamma_0\gamma_5)(-1)^t\psi=
				i\Psi_h \gamma_0\psi (-1)^t
$.
Oscillating signals for the positive parity states are present
in the correlation functions for the $D_s$ and $D_s^*$ mesons,
but they are small and hard to
see, as can be seen in fig.~\ref{fig:plotg}, 
Fig.~\ref{fig:plotf} shows the same correlation function with
the leading exponential scaled out.
The decay of the oscillating signal gives the mass of the $0^+$ state.
Because it is an excited state, the presence of a plateau is more
difficult to ascertain than for a ground state.

\begin{figure}[htb]
\vspace{9pt}
\includegraphics[width=75mm]{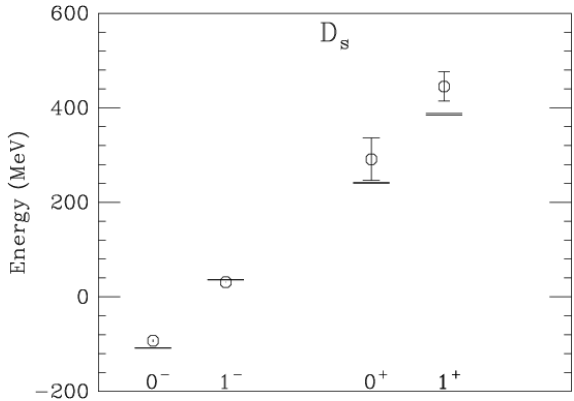}
\caption{The $D_s$   spectrum on the $m_l,m_s = 0.01,0.05$ lattices.
}
\label{fig:0105}
\end{figure}

\section{SPECTRUM}

The $D_s$ spectrum on the $m_l,m_s=0.01,0.05$ lattices is shown in
fig.~\ref{fig:0105}.
Very little dependence on the sea quark mass is observed.
The $0^+$ and $1^+$ states lie somewhat above experiment.
However, they are excited states and are somewhat more sensitive to
the choices of the priors for the higher states in the Bayes fits
than are the ground  states.
We obtain 0.86 (2) for the hyperfine splitting divided by its
experimental value.
This may be compared with 0.82 (2) for the hyperfine splitting in
charmonium.
The most likely source for these errors is the one-loop correction
to $O(a)$ $\overline{\psi}\Sigma \cdot B\psi$ operator in the clover
action.  Since this operator contributes twice to the charmonium
hyperfine splitting and once to the the $D_s$ hyperfine splitting, it is
reasonable to expect a larger effect in charmonium.

\section{LEPTONIC DECAY AMPLITUDE}

\begin{figure}[htb]
\vspace{9pt}
\includegraphics[width=65mm]{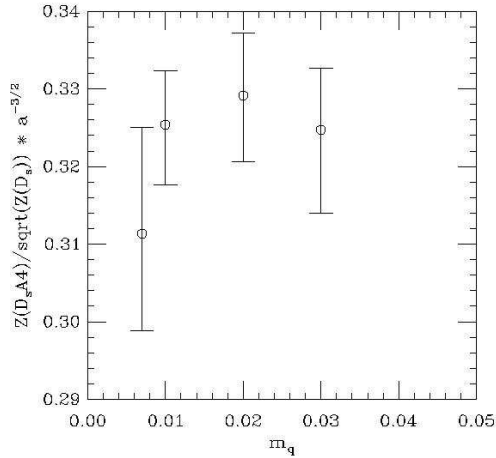}
\caption{The $D_s$ leptonic decay amplitude as a function of
sea quark mass.
}
\label{fig:fDs}
\end{figure}

This work is part of a larger project with the MILC
collaboration.
Preliminary results for the amplitude for leptonic decay of the $D_s$
are shown in fig.~\ref{fig:fDs} as a function of the sea
quark mass (using $m_s=0.041$).
There is little dependence on the sea quark mass, as expected.
The one-loop Fermilab heavy-staggered light axial current 
renormalization is in progress \cite{Nob03}.
To make an estimate of the one-loop 
correction, we use the formula 
$Z^{hl}_A= \rho_A \sqrt{Z^{hh}_V Z^{ll}_V} $~\cite{Harada:2002jh},
using 
$Z^{hh}_V=1.33(2)$ and
$Z^{ll}_V=0.86(5)$ \cite{Okamoto:2003ur}.
This leads to a current result for the decay constant of
$ f_{D_s} = 240$ MeV +/- ${\cal O}(\alpha)$.
However the order $\alpha$ correction could be large,
potentially as large as 30\%.
This is to be compared with Ryan's unquenched world average of
250 (30) MeV \cite{Ryan:2001ej}
and the unquenched NRQCD heavy staggered light result of
289($\alpha^2$) \cite{Wingate:2003ni}.
The CKM independent quantity 
$f_{D_s}/f_+^{D\rightarrow K}$
can be formed by combining this result with the result for the
semileptonic amplitude in ref.~\cite{Okamoto:2003ur}.
Since high precision is the ultimate goal, 
we will not quote a result in this early stage of the error analysis.

\section*{ACKNOWLEGEMENTS}
We thank
Claude Bernard, Christine Davies, Junko Shigemitsu, Peter Lepage, 
and Matt Wingate for helpful conversations.
We thank the MILC Collaboration for their gauge configurations.
These calculations were done on the PC clusters deployed at
Fermilab under the Doe SciDAC Program.

\end{document}